# A Case Study of Spreadsheet Use within the Finance and Academic Registry units within a Higher Education Institution

Simon Thorne and Jamie Hancock

Cardiff Metropolitan University
sthorne@cardiffmet.ac.uk

**Abstract**

This paper presents the findings of a case study of spreadsheet use in a higher education institution in the UK. The paper considers the use of spreadsheets in two units of the organisation, academic registry and finance. Spreadsheet use is explored in terms of importance, training, experience, purpose, techniques deployed, size of spreadsheets created and sharing of spreadsheets. The implications of the results are then considered in terms of accurate reporting to external funding bodies such the funding councils, internal data integrity and internal data efficiencies. The results show a large volume of spreadsheets being created and used, that the profile of spreadsheet developers is typical of other studies of spreadsheet use and the need for the organisation to have clear principles and guidelines for the development of spreadsheet models in the organisation to ensure data integrity, reduce duplication of effort and to optimise the use of spreadsheets to meet the institutions goals.

## 1.0 Introduction

This paper is generated from a case study of spreadsheet use in a higher education institution focusing on two departments that were thought to have a potentially high concentration of spreadsheets. The institution sanctioned the research in an effort to better understand and control spreadsheet use with the organisation. This report is comprised of data generated through a detailed questionnaire, informal unstructured interviews and some background investigation of spreadsheets stored on internal shared network drives.

### 1.1 Aim and research questions

The aim of this research is to appraise and quantify the veracity by which spreadsheets are used in the institution and identify the risks associated with such use.

The study was conducted in two units within a higher education institution, the academic registry and finance units since they were deemed to have the highest concentrations of spreadsheets.

### 1.2 Research questions

- How critical are spreadsheets to the units studied?
- What approaches to standards, validity and development are undertaken by institution staff using spreadsheets
- Is it possible to utilise existing corporate systems instead of using spreadsheets?
- Could the institution educate its staff to use spreadsheets more efficiently?

## 2.0 Method

The two units selected for the study were academic registry and the finance unit. These two units were selected because they were likely to have the highest concentrations of spreadsheet developers and users.

## 2.1 Academic Registry and Finance Units

Academic Registry (AR) is a sub unit of the Institution that is concerned with official record keeping and student data, communicating with external funding bodies and regulators and producing statistics used by external bodies. Registry employs 35 full time staff across various different roles all concerned with official record keeping of various types.

Recording keeping and student data is fundamental to how the institution keeps track of student progress, fees, degree awards and other relevant student related activities. The unit communicates with external funding bodies such as the Higher Education Funding Council (HEFCE) who ensure the institution compliance on fees and academic standards. The unit produces attainment statistics used for various purposes but are used in part for the universities league table calculations.

The Finance Unit (FU) is concerned with internal budgeting, student fee collection, forecasting and management of the universities finances. Internal budgeting processes manage the operational running costs of the institution, collecting student fees, managing debtors and forecasting the universities future developments. The finance department consists of 15 full time members of staff.

## 2.2 Research process

Two separate processes were undertaken as part of this study. Firstly a data capture phase sought out all spreadsheet files saved onto shared drives within the finance and academic registry units. Ideally, this would have been extended to personal hard disks on desktop machines but this was not possible through privacy concerns. The intention was not to audit the spreadsheets found rather to generate an overall impression of the scale of spreadsheet use in institution and these two units.

The second part of the study explores a questionnaire designed to probe the activities and attitudes of those working in the registry and finance units. This questionnaire is a modified version of Baker *et al.* (2008) and as such is designed to probe processes, approaches and standards of spreadsheet creation within the departments. The questionnaire was circulated to all 50 members of both units with 39 responses returned. Some questions were only partially responded to and some questions allow more than one response.

The questionnaire asked questions around the following themes:

- Levels of spreadsheet usage.
- Approaches to spreadsheet creation.
- Training acquired and available.
- Organisation standards and policies.
- Awareness of risk

## 3.0 Results

This section considers the data that has been produced from the questionnaires sent to the Finance and the Academic Registry Units. This chapter will interpret the data to show any trends and relationships which can be attributed to spreadsheet risk.

## 3.1 An overview of spreadsheets in the institution

As part of the study, a script was written to scan all shared drives across the entire institution to discover the sheer number of spreadsheets in use or stored on servers. The results showed a staggering 228,704 spreadsheets on shared drives for around 1000 staff in the institution, this number excludes spreadsheets sitting on personal hard disks. For the units of study, AR had a total 39,006 spreadsheets and FU had 39,311. Of these 78,317 spreadsheets in both AR and FU departments, it is likely that a large percentage of these are not actively used. However, due to scale it was not possible to determine with any accuracy what these figures are. What can be concluded though is that the institution is paying to store a large number of spreadsheets that probably have no current application, that could be deleted or archived and could possibly present a significant GDPR threat to the institution.

### 3.2 Questionnaire Responses

It could have been possible to have upwards of sixty responses to the questionnaires, but due to factors such as internal audits, sick leave and holidays the total number of responses was thirty nine. That number consisted of twenty-four from Academic Registry (from a potential of 35) and fifteen from Finance, which represents the whole department.

### 3.2.1 Experience and Training

The results of the first question in this section showed that participants from both units use Microsoft Excel. A very low percentage use similar programs in conjunction with Excel.

| Spreadsheet development experience levels | AR | FU | Total | % |
|---|---|---|---|---|
| Little or no experience | 6 | 5 | 11 | 29% |
| Some experience | 11 | 5 | 16 | 42% |
| Extensive experience | 3 | 4 | 7 | 18% |
| Expert experience | 3 | 1 | 4 | 11% |

*Table 1 Spreadsheet development experience levels*

Table 1 shows that the large majority (71%) have either 'little or no experience' or 'some' experience. These findings are then reflected in table 2 which shows only 27% as having formal training and 52% as having either no or colleague demonstration training.

| Sources of spreadsheet training | AR | FU | Total | % |
|---|---|---|---|---|
| None | 8 | 2 | 10 | 18% |
| Formal classroom instruction | 7 | 8 | 15 | 27% |
| Occasional informal training | 4 | 2 | 6 | 11% |
| Book based | 4 | 2 | 6 | 11% |
| Colleague demonstrations | 11 | 8 | 19 | 34% |

*Table 2 Sources of spreadsheet training (multiple answer question)*

### 3.2.2 Purpose, Functions and Use

Table 3 shows the main purposes of spreadsheets created by the respondents. The highest proportion of the sample indicated they use spreadsheets for data analysis (finance and operational) with maintaining lists and tracking data being the second most popular choice.

Breaking the results down into the different departments, FU, as expected, predominantly use spreadsheets to analyse data, 80%, with tracking data, such as budgets, being second, 60%. Over half of staff in the FU use statistical analysis, 55, see table 4. Budgets are critical to the financial stability of institution, so this enforces the finding that Finance consider spreadsheets critical to their role. AR use spreadsheets for three predominate reasons, maintaining lists 64%, tracking data 50% and analysing data 50%, see table 4.

| Spreadsheet modelling purposes | AR | FU | Total | % |
|---|---|---|---|---|
| Maintaining lists e.g. names and addresses | 14 | 6 | 20 | 27% |
| Tracking data e.g. Budgets, Sales, Inventories | 11 | 9 | 20 | 27% |
| Data analysis e.g. financial, operational | 11 | 12 | 23 | 31% |
| Determining trends and projections | 3 | 3 | 6 | 8% |
| Evaluating alternatives | 1 | 3 | 4 | 5% |
| Other | 2 | 0 | 2 | 3% |

*Table 3 Spreadsheet modelling purposes (Multiple answer question)*

In both the AR and FU responses (table 3), data analysis implies non-trivial spreadsheet programming whilst tracking data and maintaining lists is probably less computationally complex.

Table 4 shows the functions used in spreadsheets, as can be seen "none of the above" is the most popular choice. This is mostly down to the FU department who employ 'financial instruments' in their spreadsheets but don't consider those instruments to fit within the specified categories. In the AR department, statistical analysis was the most popular technique, which ties in with the nature of the work of the AR department.

| Functions used in spreadsheets | AR | FU | Total | % |
|---|---|---|---|---|
| Statistical analysis | 6 | 6 | 12 | 39% |
| Optimisation | 0 | 1 | 1 | 3% |
| Simulation | 1 | 2 | 3 | 10% |
| None of the above | 10 | 5 | 15 | 48% |

*Table 4 Functions used in spreadsheets*

When asked how many spreadsheets are used per week, 64%, use between 2 and 5 different spreadsheets per week, see table 5

| Spreadsheets used per week | AR | FU | Total | % |
|---|---|---|---|---|
| 0 | 3 | 2 | 5 | 14% |
| 1 | 3 | 2 | 5 | 14% |
| 2 to 5 | 17 | 6 | 23 | 64% |
| Over 10 | 1 | 2 | 3 | 8% |

*Table 5 Spreadsheets used per week*

When asked if the participants create spreadsheets, the majority of participants create spreadsheets as part of their job, FU 87% and AR 91%. When asked if the spreadsheets created are used to inform their line management, 40% of FU staff and only 9% of AR staff indicated so.

### 3.2.3 Spreadsheet Creation

When asked about the first steps in creating a spreadsheet, 60% of participants from the FU and 70% from AR say that they only sometimes create spreadsheets from scratch. This implies that templates or previous iterations of spreadsheets are used as a starting point for both AR and FU departments. If there are mistakes in these previous iterations, then these mistakes are compounded every time the spreadsheet is used as a starting point.

When asked about the size of the spreadsheets created the majority of participants from the FU create spreadsheets with more cells than AR, 43% create spreadsheet between 101 – 10,000 cells. Interestingly, Finance have two spreadsheets and AR have one spreadsheet that has between 10,001 – 100,000 cells.

When asked about the first step the participants take in creating a spreadsheet, 47% of staff input data directly into the spreadsheet without any planning, 42% indicated that they borrow a design from an existing spreadsheet and only 7% sketch the idea before engaging the software.

| First steps in creating a spreadsheet | AR | FU | Total | % |
|---|---|---|---|---|
| Borrow a design | 9 | 7 | 16 | 42% |
| Sketch the spreadsheet on paper | 2 | 1 | 3 | 7% |
| Enter data directly | 12 | 6 | 18 | 47% |
| Other | 0 | 1 | 1 | 2% |

*Table 6 First steps in creating a spreadsheet*

Table 7 shows further evidence of inadequate preparation. No one in the FU and very few staff in AR use any form of design methodology. This is obviously a risky practice, especially considering that some spreadsheets are very large with more than 10,000 cells.

| Do you use a software development methodology? | AR | FU | Total | % |
|---|---|---|---|---|
| Always | 0 | 1 | 1 | 3% |
| Sometimes | 0 | 1 | 1 | 3% |
| Never | 14 | 21 | 35 | 95% |

*Table 7 Do you use a software development methodology for creating spreadsheets?*

### 3.2.4 Testing and documentation

When asked about testing spreadsheet models, 59% said they never test and only 11% said they always test, see table 8.

| Do you test the spreadsheets you create? | AR | FU | Total | % |
|---|---|---|---|---|
| Never | 15 | 7 | 22 | 59% |
| Sometimes | 4 | 2 | 6 | 16% |
| Usually | 2 | 3 | 5 | 14% |
| Always | 1 | 3 | 4 | 11% |

*Table 8 Do you test the spreadsheets you create?*

Both Units have a high percentage of staff that do not test the spreadsheets they create (table 8), those that do test their spreadsheets, most use their own "common sense", see table 9.

| Approaches to testing spreadsheets | AR | FU | Total | % |
|---|---|---|---|---|
| Test extreme cases | 2 | 1 | 3 | 7% |
| Manually calculate answers | 3 | 4 | 7 | 16% |
| Display formulae | 2 | 5 | 7 | 16% |
| Code inspect formulae | 3 | 5 | 8 | 18% |
| Test performance for plausibility | 2 | 3 | 5 | 11% |
| Error checking in Excel | 1 | 1 | 2 | 4% |
| Use Common sense | 6 | 7 | 13 | 29% |

*Table 9 Approaches to testing spreadsheets (Multiple answer question)*

Table 10 shows the majority of staff (58%) do not document any of their spreadsheets. Of those who said they do document, they most commonly spend only 1-10% of the development time on documenting the spreadsheets they create.

| Do you document your spreadsheets? | AR | FU | Total | % |
| --- | --- | --- | --- | --- |
| Never | 11 | 10 | 21 | 58% |
| Sometimes | 10 | 3 | 13 | 36% |
| Usually | 1 | 1 | 2 | 6% |

*Table 10 Do you document your spreadsheets?*

### 3.2.5 Dissemination and sharing

64% of FU and AR staff share their spreadsheets with between 2 and 5 people, as shown in table 10.

| Typical number of users per spreadsheet | AR | FU | Total | % |
| --- | --- | --- | --- | --- |
| 0 | 3 | 2 | 5 | 14% |
| 1 | 4 | 1 | 5 | 14% |
| 2 to 5 | 12 | 11 | 23 | 62% |
| 6-10 | 0 | 1 | 1 | 2% |
| Over 10 | 2 | 1 | 3 | 8% |

*Table 11 Typical number of users per spreadsheet*

As shown in table 12 the majority of staff from both Units share their entire spreadsheet model via email or sharepoint.

| Approaches to sharing spreadsheets | AR | FU | Total | % |
| --- | --- | --- | --- | --- |
| I rarely share any part of my spreadsheets | 5 | 4 | 9 | 19% |
| I provide a summary of results | 4 | 3 | 7 | 15% |
| I provide parts of my spreadsheets | 2 | 3 | 5 | 11% |
| I share the entire model | 16 | 10 | 26 | 55% |

*Table 12 Approaches to sharing spreadsheets*

### 3.2.6 Modification and backup

With more than 39,000 spreadsheets stored on the network by both units, it was surprising to find that participants from the Finance Unit commented that only 29% of major spreadsheets have a lifetime of more than 2 years and in the AR only 9%. Both AR and FU responses are shown in table 13. This suggests that the vast majority of spreadsheets stored on the institutions servers are redundant. The disk space these files are using is significant and could be utilised more efficiently.

| My organisations uses standards for spreadsheet development | AR | FU | Total | % |
| --- | --- | --- | --- | --- |
| One week or less | 3 | 1 | 4 | 11% |
| A few weeks or months | 8 | 6 | 14 | 38% |
| One to two years | 10 | 3 | 13 | 35% |
| More than two years | 2 | 4 | 6 | 16% |

*Table 13 what is the typical lifecycle of the spreadsheets you create?*

When asked about backing up of spreadsheet models, about an equal amount use sharepoint or personal storage space provided by the institution.

| Backup destinations | AR | FU | Total | % |
|---|---|---|---|---|
| No backup, save to C: | 4 | 7 | 11 | 19% |
| USB Flash drive | 4 | 1 | 5 | 9% |
| Sharepoint | 14 | 8 | 22 | 38% |
| Networked personal space | 18 | 2 | 20 | 34% |

*Table 14 Backup destinations*

When asked how frequently participants archive their spreadsheets, the response showed that very few archive regularly.

| How frequently do you archive your spreadsheets | AR | FU | Total | % |
|---|---|---|---|---|
| Rarely if ever | 7 | 10 | 17 | 47% |
| Occasionally | 11 | 3 | 14 | 39% |
| Frequently | 3 | 2 | 5 | 14% |

*Table 15 How frequently do you archive your spreadsheets?*

Considering the data in table 15, it is possible to understand why there are so many spreadsheets (39,000) in comparison to the amount of staff members. The results of the question show that only 14% frequently archive their work with 47% saying they rarely if ever archive. This is part of the reason why so many 'active' spreadsheets exist on shared drives within these departments.

### 3.2.7 Standards, Policies and Risk Perception

The responses from the participants in both Units suggest they know very little, if anything regarding spreadsheet standards and policies employed by the institution. The question asked about standards the institution sets, as shown in Table 16, 76% answered 'I don't know' and 13% declaring there are no standards.

| My organisations uses standards for spreadsheet development? | AR | FU | Total | % |
|---|---|---|---|---|
| No Standards | 3 | 2 | 5 | 13% |
| Informal guidelines | 2 | 1 | 3 | 8% |
| Formal written guidelines | 0 | 1 | 1 | 3% |
| I don't know | 18 | 11 | 29 | 76% |

*Table 16 My organisations uses standards for spreadsheet development*

When asked about how important spreadsheets are to the institution as a whole, 60% said very important or critical which shows that the participants perceive the use of spreadsheets to be central to the success of the institution, see table 17.

| How important are spreadsheets to the institution? | AR | FU | Total | % |
|---|---|---|---|---|
| Moderately important | 8 | 6 | 14 | 40% |
| Very important | 9 | 1 | 10 | 28% |
| Critical | 5 | 6 | 11 | 32% |

*Table 17 How important are spreadsheets to the institution?*

When asked how critical spreadsheets were to the specific unit the participants worked in, similar answers to table 17 were gathered.

| How important are spreadsheets to your individual unit? | AR | FU | Total | % |
|---|---|---|---|---|
| Moderately important | 6 | 4 | 10 | 26% |
| Very important | 12 | 3 | 15 | 39% |
| Critical | 5 | 8 | 13 | 34% |

*Table 18 How important are spreadsheets to your individual unit?*

From tables 17 and 18 it is clear that the participants understand the criticality of spreadsheets to the institution with 60% indicating they thought spreadsheets were either critical or very important to the institution, see table 17, and 73% indicating they were critical or very important the unit the participants worked in, see table 18. If we consider table 1, which was concerned with training and experience, we can conclude that even though 73% consider spreadsheets as very important or critical to their unit, 30% consider themselves to have little or no experience and only 27% have had any formal training.

When asked about the risks that spreadsheets pose to the institution, 74% of participants indicated that they thought spreadsheets posed a medium or high risk, see table 19. In combination with tables 17 and 18, it would seem that the participants have a good understanding of the level of risks posed by spreadsheets given their criticality to the institution.

| Perceived level of risk spreadsheets pose | AR | FU | Total | % |
|---|---|---|---|---|
| High risk | 2 | 2 | 4 | 11% |
| Medium risk | 15 | 7 | 22 | 63% |
| Low risk | 3 | 2 | 5 | 14% |
| No risk | 2 | 2 | 4 | 11% |

*Table 19 Perceived level of risk spreadsheets pose to the institution*

When asked about who should manage the risks spreadsheets pose, 86% indicated they didn't know who was responsible and only 5 and 8% thought it was the managers or the developers responsibility respectively, see table 20.

| Who is responsible for managing the risks spreadsheets pose? | AR | FU | Total | % |
|---|---|---|---|---|
| The developer | 2 | 1 | 3 | 8% |
| The manager | 1 | 1 | 2 | 5% |
| Don't know | 20 | 12 | 32 | 86% |

*Table 20 Who is responsible for managing the risks spreadsheets pose?*

This confusion is likely down to a failure to develop and communicate a clear policy around spreadsheet across the institution, the answers here are mirrored in table 16 which shows that participants overwhelmingly "didn't know" if there were policies and standards for spreadsheet use at the institution.

These results show that whilst participants are aware of some of the risks spreadsheets pose to the institution, they have no idea how to mitigate these risks, what the standards are they should adhere to and who bears responsibility for these risks. A large part of this failing comes directly from the centre of the institution who must be unaware of the risks posed by spreadsheets to institution.

### 3.3 Summary of risks to the institution

Given the data collected, the institution is placing themselves at risk in a number of veins. There is the risk of making errors in spreadsheets that go unnoticed and propagate to strategic decision making. There are risks in the institution communicating erroneous data to third parties and regulators via data entry or erroneous analysis in spreadsheets. There is a risk of fraudulent activities going unnoticed through lack of oversight, ownership and standards governing spreadsheets. From the FU, there is the risk financial planning, forecasting and tracking could be compromised resulting in internal failure to plan and budget, there could be dire financial consequences as a direct result. For AR which is more student focussed, risks exist in the mishandling of data, errors in student data and errors in reporting to external bodies which could put the institution in legal and financial jeopardy.

### 3.4 Criticality

From the evidence available, spreadsheets play a critical role in the institution on a number of levels. All financial planning and reporting is conducted using spreadsheets as evidenced by the FU responses to the survey. Information is collected, analysed and communicated to external bodies such as the Higher Education Statistics Authority (HESA) and other regulatory bodies that can affect levels of funding to the institution, strategic choices for the institution's direction and how the external credibility of the institution is viewed.

If we apply criticality tests developed by Hamill and Chambers (2008), McGeady and McGouran (2008) or Thorne and Shubbak (2015), a large number of spreadsheet applications would be deemed critical and in need of management. If we apply other criticality tests such as those used to determine Sarbanes Oxley (SOX) compliance, communicating with external bodies and internal financial planning would place a number of spreadsheet applications in the critical category, which would need formal management.

Hence, the institution has a significant number of critical spreadsheet applications that need proper mitigation. The data obtained in this survey suggests there are a number of areas that need immediate and focussed attention to reduce the likelihood of a major problem arising.

### 3.5 Specific Spreadsheet Risks

As with any organisation using spreadsheets, there is a risk of making errors in critical spreadsheet applications. These errors risk the integrity of the internal decision making process but the institution also faces risks arising from other practices too such as communicating with third parties, fraud and falsification and GDPR.

The validity of the models built could be strengthened by addressing the weaknesses identified in documentation, complexity, training, organisational standards, development methodologies and testing.

### 3.5.1 Documentation

Table 10 shows the majority of staff from both Units do not document any of their spreadsheets. Of those who said they do document, they most commonly spend only 1-10% of the development time on documenting the spreadsheets they create. Documentation is vital for the future maintenance of spreadsheets (Pryor, 2006), since most indicate they do not document, there is a significant risk that an undocumented spreadsheet may become a major problem for the organisation if a key employee leaves without imparting crucial information.

Cleary *et al.* (2003) discuss one such example where a spreadsheet was used to manage contracts at a large NHS health trust. The employee responsible for the spreadsheet left the organisation without providing any documentation on the function and purpose of the spreadsheet. This spreadsheet managed clinical contracts, it took data from corporate databases, manipulated the data inside the spreadsheet and then inserted data back into the corporate database bypassing all validation and integrity controls on the database. Since its function was critical to the health trust and it was a 'blackbox', it could not be simply removed but since there was no documentation, it could not be maintained or checked for validity. Eventually the trust engaged a software house to forensically dismantle the spreadsheet application and engineer software to replace it.

### 3.5.2 Complexity of spreadsheet tasks

In terms of the complexity of spreadsheet modelling activities, figure 3 implies that some of the activities carried out are non-trivial spreadsheet programming tasks that include writing formulae which are considered more complex than simple data entry and tracking tasks.

'Complex' spreadsheet modelling activities generally attract higher error rates of around 5% Base Error Rate (BER). Maintaining lists and tracking data attracts a smaller BER of around 2.5% in data entry (Panko, 2008).

Hence, there is a significant chance that around 5% of the spreadsheets used for data analysis contain material errors and 2.5% of the data used for maintaining lists and tracking is erroneous. In addition, if the data used in maintaining lists and tracking is of a personal nature, there is a risk that the institution is exposing themselves to GDPR risks especially if the spreadsheets in question are not active and dormant. Indeed this must be a risk for many organisations who possess dormant spreadsheets containing personal information.

In terms of the size of spreadsheet models in the institution, the FU have two spreadsheets and AR have one spreadsheet that has between 10,001 – 100,000 cells. These large spreadsheets are probably complex, mission critical and candidates for migration to the IT function. A lack of development methods, testing, documentation and validation is a real problem when considering the scale and likely complexity and importance of 10-100K cell spreadsheets.

For the more complex applications, some mitigating retroactive action should be taken, such as code inspection. It may also be prudent to consider migration of some of these applications to formally designed software led by the information systems department inside the organisation.

### 3.5.3 Formal training

Tables 1 and 2 show that the majority of staff have no formal training and cite their own experience levels as either beginners with some experience or no real experience. These findings are typical and reflected in other studies, which show that spreadsheet developers usually have no formal training and little experience (Taylor *et al.* 1998, Grossman and Ozluk 2004 Panko 2008, Baker *et al.*2006, Thorne and Altarawneh 2017, Thorne and Shubbak 2015, Mireault and Gresham 2015). Hence, the profile of the participants is broadly typical of other spreadsheet users worldwide.

The institution should aim to provide some basic how to type training and provide some training on the context of risk in spreadsheet development as advocated by Hamill and Chambers (2008).

### 3.5.4 Organisational standards

Table 16 shows the majority of respondents 'don't know' if there are standards surrounding the development and deployment of spreadsheets.

The institution does not have any such policy on the development, management and maintenance of spreadsheet applications, which explains the responses given by participants. One participant indicated that there were formal written standards but this participant must have been mistaken since no guidelines exist of this nature. The institution should develop guidelines based around an assessment of criticality of each individual spreadsheet application to ensure that sufficiently critical applications are subject to some minimum standards.

### 3.5.5 Development methodologies

Tables 6 and 7 show that approaches to development are ad-hoc with most participants entering data directly into the spreadsheet or borrowing a pre-existing design. Participants also indicated that they don't use any sort of planning methodology for developing their spreadsheets.

The institution should impose methodology standards on spreadsheet applications that are considered critical; many papers discuss how the use of a methodology can help improve the quality control and the accuracy of the spreadsheet. (Butler 2000, Diemer 2002, Thorne 2009, Thorne and Altarawneh 2017, Thorne and Shubbak 2015, Mireault and Gresham 2015)

### 3.5.7 Lack of testing

Tables 8 and 9 show participant approaches to testing with 59% saying they do not test their applications. For those that say they do test, most cite 'common sense' as the means for testing.

It is likely that the common sense test is the same as the "sniff test" discussed by Grossman *et al.* (2010) and Caulkins *et al.* (2007) which is essentially a visual check to see if the numerical values are approximately correct. In other words, these respondents do not test their spreadsheet models with any rigour or reliability. A few do employ some techniques such as code inspection, manually calculating the outcomes, plausibility testing and error checking functions in excel. However, we can infer the large majority of these models are not tested rigorously or reliably. Even those who do audit their own work are still bound by the probability that alone they will find around 60% of their own mistakes (Panko, 2008).

### 3.5.8 Communicating with third parties

The institution places themselves at particular risk in communicating data or analysis conducted in spreadsheets with external third parties. This is especially so if those third parties are regulatory, since making erroneous statements or having to retract statements would cause the institution embarrassment, damage its reputation and could have financial or legal implications. The institution engages with a number of regulatory bodies, both the FU and AR units provide statistics, forecasts and accounting for the Higher Education Statistics Authority and the Funding Council. Retracting financial or statistical statements could have knock on effect on the institution in terms of funding, reputation and legal consequences. In the case of AR, mistakes in student records or statistics surrounding students could be pursued legally by those affected (Nightingale, 2017).

### 3.5.9 Fraud and falsification

The institution also risks fraud or data falsification incidents through a basic lack of oversight and standards. Spreadsheets are often the vehicle used to commit fraud since they can be used to hide fraudulent transactions and create a clean falsified version of events. Thorne (2013) discusses examples of spreadsheet fraud and falsification involving spreadsheets in finance. Spreadsheets can be used to create false paper trails, making fraudulent transactions seem legitimate. The mechanism of falsification can be difficult to spot inside spreadsheets too, for the same reason genuine errors are hard to find. The mix of data and programming structures and the ability to hide parts of the calculation within the software or away from the visible portion of the spreadsheet makes spotting fraudulent calculations difficult. One approach to securing artefacts critical to the organisation would be to adopt a similar approach to the pharmaceutical industry who have electronic vaults that contain spreadsheets that present drug control trials.

Title 21 CFR Part 11(FDA, 2011) protects against security violations and ensures the reliability of electronic records via the use of digital vaults to store official record evidence base for the effectiveness of drugs  Spreadsheets are explicitly mentioned in this legislation which demands companies provide evidence of: audits, validation, electronic signatures and documentation for any software artefact including spreadsheets. The legislation also dictates that electronic artefacts such as spreadsheets be stored in a secure server so that once the spreadsheet has been created and audited, it cannot be changed without authorisation. This should discourage fraud since the process of auditing, validating and documentation would most likely uncover any mistakes or fraud in spreadsheets. Such a system would prevent fraudulent alteration of spreadsheets, since access is limited by 'lock and key'. Authorisation would be needed to implement any changes with subsequent auditing, validation and documentation. Such a system would make the prospect of committing fraud fraught with difficulties, higher risk and therefore a less attractive option. In addition if all organisations relying on spreadsheets were to use a system that complies with Title 21 CFR Part 11, regulators could easily check compliance.

### 3.5.10 General Data Protection Regulations (GDPR)

The institution also faces risks arising from GDPR introduced in the EU in 2018. This legislation dictates how information can be collected, stored, pseudo- anonymised processed and disposed of for all EU citizens. Any data that can personally identify an individual is a potential GDPR breach if such information were accidentally imparted to an unintended party. GDPR breaches are usually dealt with via fines that can amount to €10 Million or 2% of the total turnover. Breaches can arise in any of the following areas: Implementation of privacy by design, data processing activities, recording of data processing activities, the data processor's main obligations, notifications of breaches and appointments of data protection officers.

A fine of up to €20 million or 4% of total turnover for breaches in basic data processing including privacy consent, individuals' rights of access and to be forgotten and transfer of data outside of the European Economic Area. The AR and FU units have some 39,000 active spreadsheets, some of which must contain personal identifiable data. Particularly in the AR unit, there are most likely a substantial number of spreadsheets that contain information that can personally identify students inside the institution, these spreadsheets if accidentally imparted could cost the institution a large amount of money. The Italian Data Protection Authority recently issued fines amounting to €11 Million against five separate companies for breaches in transferring information outside of the EU (Corragio, 2019). With most higher education institutions operating on a truly international scale, it is not infeasible a similar situation could arise through the transfer of information internationally at the institution.

Beyond the AR and FU units, the institution stores approximately another 190,000 spreadsheets, some of which are very likely to contain personal identifiable data. It would be reasonable to assume that since the FU and AR units generally do not archive their work, this is reflected across the institution and that a good number of these spreadsheets are no longer needed for an active purpose. Hence, the institutions risk of GDPR breaches is significant and immediate action to pseudo-anonymise or delete the information should be taken.

### 3.6 Mitigation of risks at the institution

The following practical steps should be taken over the short and long term to address the risks the institution places themselves at

1. **Generate an inventory of active critical spreadsheet applications and take appropriate mitigating actions with the riskiest spreadsheets.**

This step is critical to controlling the institutions spreadsheet risk and should identify the highest risk applications and take immediate action to secure them. This should at the very least allow the institution oversight on the critical applications currently in use and provide a Key Risk Indicator (KRI) that the institution can manage and track their spreadsheet risk by. One tool for assessing risk is discussed in Thorne and Shubbak (2015) called 'Risk Calculator' which can take a large volume of spreadsheets and give a relative and absolute risk score for each spreadsheet. This should identify the priority targets for mitigation activities. Mitigation activities should be matched to the risk each spreadsheet poses and should be considered a longer-term goal of the process. Hamil and Chambers (2008) discuss how mitigation activities should be applied to spreadsheets, they assert that for critical applications a number of minimum controls should be implemented: *version control, access control, change control, business continuity measures, documentation and testing*. These measures should be implemented by the institution as soon as possible. In addition, large complex spreadsheet applications should be considered for redevelopment by the IT function of the institution – there is at least one very large complex spreadsheet in the FU unit that should be considered for redevelopment.

## 2. Development of a robust spreadsheet development policy governing approaches to development, deployment, documentation, testing and accountability

The institution should immediately construct an organisation wide policy, which covers the development, deployment, documentation, testing and accountability of individual spreadsheet applications based on the criticality of individual applications. This policy should aim to make individuals and managers accountable for the quality, truthfulness and accuracy of spreadsheet artefacts in use in the organisation. This should prevent the risk the institution faces increasing in an uncontrolled manner and provides an avenue of accountability for these actions. This is essential in controlling and protecting the institution.

## 3. As part of the organisational policy, decide on development methodologies, documentation standards and testing policies for spreadsheets

The institution should adopt a structured approach to spreadsheet development that is sympathetic to demands of spreadsheet development. Grossman and Ozluk (2004) present a structured framework for spreadsheet development that takes into account the typical skills reported by spreadsheet developers. There are other approaches that would be suitable also, Mireault 2015 describes a development technique called Structured Spreadsheet Modelling and Implementation (SSMI) methodology which is designed for spreadsheet modellers. This technique is also designed with non-computing professionals in mind and could be used to model and implement complex spreadsheet models.

In a similar vein, documentation standards should also be considered and adopted. Standards presented by O'Beirne (2005) and Pryor (2006) both advocate using documentation within the spreadsheet itself in some cases using cell comments to give more detail on data structures and formulae and in other cases using a welcome sheet that covers the basic principles of what the spreadsheet does and how it calculates.

For testing strategies, Panko (2006) advocates a variety of testing approaches, some from traditional software development such as unit testing and others that are more tailored to spreadsheets such as the Fagan method. Panko also discusses code inspection and auditing as a means to find and correct errors, which remains the most effective approach.

## 4. Identify and deliver appropriate training in planning, development and testing of spreadsheet models

From the results of this survey, there is a clear need to train staff in AR and FU units in the basics of spreadsheet development, use of methodologies, production of documentation and testing.

## 5. Peer working, user groups and risk awareness sessions

The institution should establish a spreadsheet user group and encourage members to work collaboratively in development and testing of spreadsheet models. This group should have it's own internal space on the network and should be allowed to share suitable spreadsheet templates, tips on writing spreadsheets and a mechanism for peer review of spreadsheet models. Chambers and Hamill (2008) and McGeady and McGouran (2008) both advocate spreadsheet user groups and noted the impact of presenting spreadsheet modellers with facts about spreadsheet risk.

### 3.7 Conclusions

In summary, the institutions approach to using spreadsheets is typical of most organisations. There is a lack of awareness of how critical spreadsheets are to the running of the institution, a lack of awareness over the pervasiveness of errors and the risks spreadsheets pose. Staff developing spreadsheets are not trained and generally have 'self taught' experience, they do not plan, develop with a methodology or test their models. The organisation has no standards or policies whatsoever to govern the use of such computing artefacts. As a direct result, the institution places itself at significant risk of a decision being made on bad numbers, assumptions or models for which the consequences could be severe. This report has highlighted those points and others in detail and provided a framework for mitigating these risks for the institution.